\documentclass{webofc}
\usepackage{amsmath,amsfonts,mathrsfs,amssymb}
\usepackage{mathptmx}       
\usepackage{helvet}         
\usepackage{courier}        
\usepackage{type1cm}        
\usepackage{makeidx}         
\usepackage{graphicx}        
\usepackage{multicol}        
\usepackage[bottom]{footmisc}
\usepackage{epsfig}

\makeindex             
\usepackage[varg]{txfonts}   

\woctitle{QCD@Work 2014}
\begin{document}
\title{Dynamical holographic QCD model}
%
%

\author{Danning Li \inst{1,2}\fnsep\thanks{\email{lidn@mail.ihep.ac.cn}}
           \and
        Mei Huang \inst{1,3}\fnsep\thanks{\email{huangm@mail.ihep.ac.cn}}}

\institute{Institute of High Energy Physics, Chinese Academy of Sciences,
             Beijing 100049, China
\and  Institute of Theoretical Physics, Chinese Academy of Sciences,
             Beijing 100190, China
\and  Theoretical Physics Center for Science Facilities, Chinese Academy of Sciences,
  Beijing 100049, China}

\abstract{%
We develop a dynamical holographic QCD model, which resembles the
renormalization group from ultraviolet (UV) to infrared (IR). The dynamical
holographic model is constructed in the graviton-dilaton-scalar framework
with the dilaton background field $\Phi$ and scalar field $X$ responsible for
the gluodynamics and chiral dynamics, respectively. We summarize our results
on hadron spectra, QCD phase transition and transport properties including
the jet quenching parameter and the shear/bulk viscosity in the framework of
the dynamical holographic QCD model.
}
\maketitle
\section{Introduction}
\label{intro}
Quantum chromodynamics (QCD)  in the infrared (IR) regime still remains as an outstanding challenge
in the formulation of QCD as a local quantum field theory.  In recent decades,  the
anti-de Sitter/conformal field theory (AdS/CFT) correspondence \cite{Maldacena:1997re,Gubser:1998bc,Witten:1998qj}
provides a revolutionary method to tackle the problem of strongly coupled gauge theories.
The gauge/gravity duality has been widely used in investigating hadron physics, strongly coupled quark gluon
plasma and condensed matter. In general, holography maps quantum field theory (QFT) in
d-dimensions to quantum gravity in (d + 1)-dimensions, and the gravitational description
becoming classical when the QFT is strongly-coupled. The extra dimension
is normally regarded as an energy scale or renormalization group (RG) flow in the QFT
\cite{Adams:2012th}.

In this talk, we introduce our recently developed dynamical holographic QCD model
\cite{DhQCD}, which resembles the renormalization group from ultraviolet (UV) to
infrared (IR). The model is constructed in the graviton-dilaton-scalar framework,
the dilaton background field $\Phi(z)$  is dual to the dimension-4 gluon operator, and
the scalar field $X(z)$  is dual to the dimension-4 gluon operator at the UV boundary.
In the IR region,  the dilaton background field and scalar field are responsible
for nonperturbative gluodynamics and chiral dynamics, respectively. The metric structure
at IR can be automatically deformed by the nonperturbative gluon condensation and chiral
condensation in the vacuum. We then summarize our results on hadron spectra, QCD phase
transitions, equation of state and transport properties in this model.

\section{The scalar glueball spectra and meson spectra}
\label{sec-1}

We construct the quenched dynamical holographic QCD model for the pure gluon system
in the graviton-dilaton framework by introducing a scalar dilaton field $\Phi(z)$ in the bulk.
The 5D graviton-dilaton coupled action in the string frame takes the form of:
\begin{equation}\label{action-graviton-dilaton}
 S_G=\frac{1}{16\pi G_5}\int
 d^5x\sqrt{g_s}e^{-2\Phi}\left(R_s+4\partial_M\Phi\partial^M\Phi-V^s_G(\Phi)\right),
\end{equation}
with $G_5$ the 5D Newton constant, $g_s$, $\Phi$ and $V_G^s$ are the 5D
metric, the dilaton field and dilaton potential in the string frame, respectively.
The metric ansatz is chosen to be
\begin{equation}\label{metric-ansatz}
ds^2=b_s^2(z)(dz^2+\eta_{\mu\nu}dx^\mu dx^\nu), ~ ~ b_s(z)\equiv e^{A_s(z)}.
\end{equation}
The dilaton field takes the form of  $\Phi(z)=\mu_G^2z^2\tanh(\mu_{G^2}^4z^2/\mu_G^2)$,
which is dual to the dimension-4 gauge invariant gluon operator ${\rm Tr} G^2 $
at UV $\Phi(z)\overset{z\rightarrow0}{\rightarrow} \mu_{G^2}^4 z^4$ to meet the
requirement of gauge invariance for gauge/gravity duality, and it has the quadratic form of
$\Phi(z)\overset{z\rightarrow\infty}{\rightarrow} \mu_G^2 z^2$ at IR to produce
linear confinement \cite{Karch:2006pv}.

\begin{figure}[!htb]
\sidecaption
\epsfxsize=6.5 cm \epsfysize=6.5 cm \epsfbox{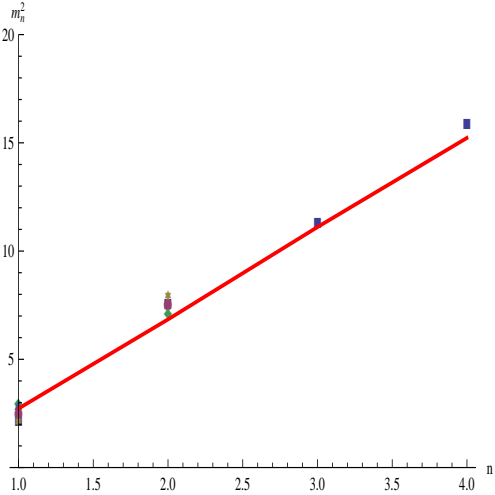}
\caption{The scalar glueball spectra in the quenched dynamical holographic QCD model with $\mu_G=1 {\rm GeV}$ and $\mu_{G^2}=1 {\rm GeV}$. }
\label{z4-z2glueball}
\end{figure}

By self-consistently solving the Einstein equations, the metric structure will be automatically
deformed at IR by the dilaton background field. The glueball can be excited from the QCD
vacuum described by the quenched dynamical holographic model,  we can solve the scalar
glueball spectra and the result is shown in Fig.\ref{z4-z2glueball},  for details, please refer to
\cite{DhQCD}. The dots are lattice data taken from \cite{glueball-lattice}. It is a surprising
result that if one self-consistently solves the metric background
under the dynamical dilaton field, it gives the correct ground state and at the same time gives the
correct Regge slope.

We add light flavors in terms of meson fields on the gluodynamical background.
The total 5D action for the graviton-dilaton-scalar system takes the  form of
\begin{equation}
 S=S_G + \frac{N_f}{N_c} S_{KKSS},
\end{equation}
with
\begin{center}
\begin{eqnarray}
 S_G=&&\frac{1}{16\pi G_5}\int
 d^5x\sqrt{g_s}e^{-2\Phi}\big(R+4\partial_M\Phi\partial^M\Phi-V_G(\Phi)\big), \\
 S_{KKSS}=&&-\int d^5x
 \sqrt{g_s}e^{-\Phi}Tr(|DX|^2+V_X(X^+X, \Phi)+\frac{1}{4g_5^2}(F_L^2+F_R^2)).
\end{eqnarray}
\end{center}

\begin{figure}[!htb]
\sidecaption
\epsfxsize=6.5 cm \epsfysize=6.5 cm \epsfbox{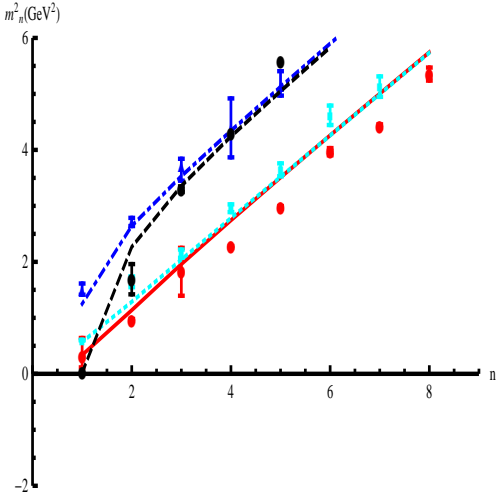}
\caption{Meson spectra in the dynamical soft-wall model  comparing with experimental data.
The red and black lines are for scalars and pseudoscalars, the green and blue
lines are for vectors and axial-vectors. }
\label{allmassespic}
\end{figure}

The meson spectra produced from this model is summarized in Fig.\ref{allmassespic}. It is observed  that in our
graviton-dilaton-scalar system,  the generated meson spectra agree well with experimental data.

\section{Phase transition and equations of state}

To study the thermodynamics, it's natural to introduce a black hole in the 5D gravity side and study the black hole thermodynamics.
The finite temperature metric ansatz in string frame becomes
\begin{equation} \label{metric-stringframe}
ds_S^2=
e^{2A_s}\left(-f(z)dt^2+\frac{dz^2}{f(z)}+dx^{i}dx^{i}\right).
\end{equation}
Under this metric ansatz, from the Einstein equations we get the following equations of motion:
\begin{center}
\begin{eqnarray}
 -A_s^{''}+A_s^{'2}+\frac{2}{3}\Phi^{''}-\frac{4}{3}A_s^{'}\Phi^{'}&=&0, \label{Eq-As-Phi-T} \\
 f''(z)+\left(3 A_s'(z) -2 \Phi '(z)\right)f'(z)&=&0,\label{Eq-As-f-T}\\
 \frac{8}{3} \partial_z
\left(e^{3A_s(z)-2\Phi} f(z)
\partial_z \Phi\right)-
e^{5A_s(z)-\frac{10}{3}\Phi}\partial_\Phi V_G^E&=&0,
\end{eqnarray}
\end{center}
with $V^E_G=e^{4\Phi/3}V_{G}^s$. At the UV region, we require the asymptotic AdS boundary condition
and we also require $\Phi$ to be finite at $z=0, z_h$ with $z_h$ the black-hole horizon where $f(z_h)=0$.
The temperature of the solution would be identified with the Hawking temperature
\begin{equation} \label{temp}
T =\frac{e^{-3A_s(z_h)+2\Phi(z_h)}}{4\pi \int_0^{z_h} e^{-3A_s(z^{\prime})+2\Phi(z^{\prime})} dz^{\prime} }.
\end{equation}
By taking the dilaton profile of $\Phi=\mu_G z^2$, the metric prefactor $A_s$ could be solved analytically as
\begin{equation}\label{As-sol}
A_s(z) =\log(\frac{L}{z})-\log(_0F_1(5/4,\frac{\mu_G^4z^4}{9}))+\frac{2}{3}\mu_G^2z^2,
\end{equation}
where $L$ is the AdS radius.

\begin{figure}[!htb]
\sidecaption
\epsfxsize=6.5 cm \epsfysize=6.5 cm \epsfbox{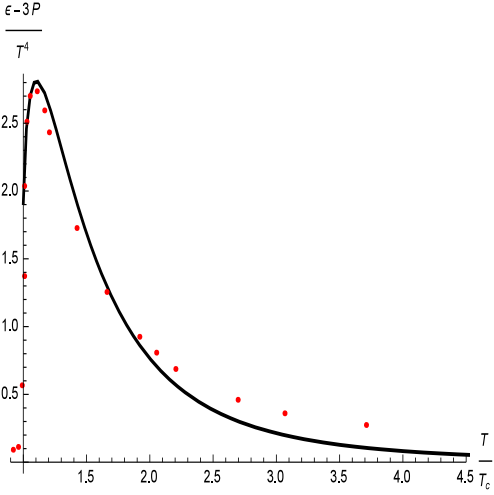}
\caption{The trace anomaly $\frac{\epsilon-3p}{T^4}$ as a function of $T_c$ scaled temperature
$T/T_c$ for $G_5=1.25$ and $\mu_G=0.75 {\rm GeV}$(Solid black line). The red dots are the
pure SU(3) lattice data from \cite{LAT-EOS-G}.} \label{Cs2-TraceA}
\end{figure}

The black-hole entropy density $s$ could be easily read from the Bekenstein-Hawking formula,
\begin{equation}
s=\frac{1}{4G_5}e^{3A_s(z_h)-2\Phi(z_h)},
\end{equation}
and the pressure density $p$, energy density $\epsilon$ can be solved from the following equations:
\begin{equation}
\frac{dp(T)}{dT}= s(T), ~~ \, \epsilon=-p+sT.
\end{equation}
We compare the result on trace anomaly $\epsilon-3p$ with the pure SU(3) lattice data from \cite{LAT-EOS-G} in
Fig.\ref{Cs2-TraceA}. It can be seen that  the thermodynamical quantities in the quenched dynamical holographic QCD
model can describe the pure gluon system quite well. The near $T_c$ sharp peak in the trace anomaly shows
that we have encoded the correct IR physics in the 5D model.

\section{Transport coefficients in the dynamical holographic QCD model}

\subsubsection{Jet quenching parameter $\hat{q}$}

\begin{figure}[!htb]
\sidecaption
\epsfxsize=7.5 cm \epsfysize=6.5 cm \epsfbox{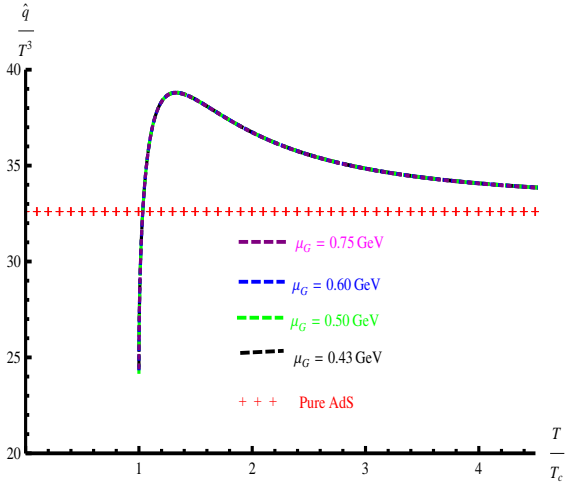}
\caption{$\hat{q/T^3}$ as a function of temperature $T$ with $G_5=1.25$ and $\lambda=6\pi$.} \label{qhatdT3}
\end{figure}

Jet quenching measures the energy loss rate of an energetic parton going thorough the created hot dense medium. In \cite{Li:2014hja}
we have had a detailed study on the temperature dependence behavior of $\hat{q}$.  Following the method in \cite{Liu:2006ug},
the jet quenching parameter is related to the adjoint Wilson loop by the following equation
\begin{equation}
W^{Adj}[\mathcal {C}]\approx exp(-\frac{1}{4\sqrt{2}}\hat{q}L^{-}L^2).
\end{equation}
In gravity side, the Wilson loop is dual to the on-shell value of the string Nambu-Goto action. Choosing proper parametrization
of string configuration (for details, please see \cite{Li:2014hja}), one can get
\begin{equation}\label{qhatfor-res}
\hat{q}=\frac{\sqrt{2}\sqrt{\lambda}}{\pi z_h^3 \int_0^{1}d\nu\sqrt{\frac{e^{-4A_s(\nu z_h)}}{z_h^4}\frac{1-f(\nu z_h)}{2}f(\nu z_h)}}.
\end{equation}
The numerical results on  $\hat{q}/T^3$ as a function of temperature is shown in Fig.\ref{qhatdT3}. It is seen that around $T=1.1 T_c$,
there's peak with the height around 40, which is totally different from the pure AdS case.

\subsubsection{Shear/bulk viscosity}

In the case of $AdS_5$, the shear viscosity over entropy density is $1/4\pi$ an the bulk viscosity is zero for a conformal system.
However, it has been observed that the shear viscosity over entropy density ratio $\eta/s$ has a
minimum in the phase transition region in systems of water, helium, nitrogen \cite{Csernai:2006zz} and
the same feature is expected to show up for QCD phase transition, and
the lattice QCD shows that the temperature dependence of bulk  viscosity over entropy density \cite{LAT-xis-KT}
exhibits a peak around the phase transition. We derive the shear/bulk viscosity in the dynamical holographic QCD model,
and the details will be given in \cite{LiHeHuang}.

\begin{figure}[h]
\begin{center}
\epsfxsize=6.5 cm \epsfysize=6.5 cm \epsfbox{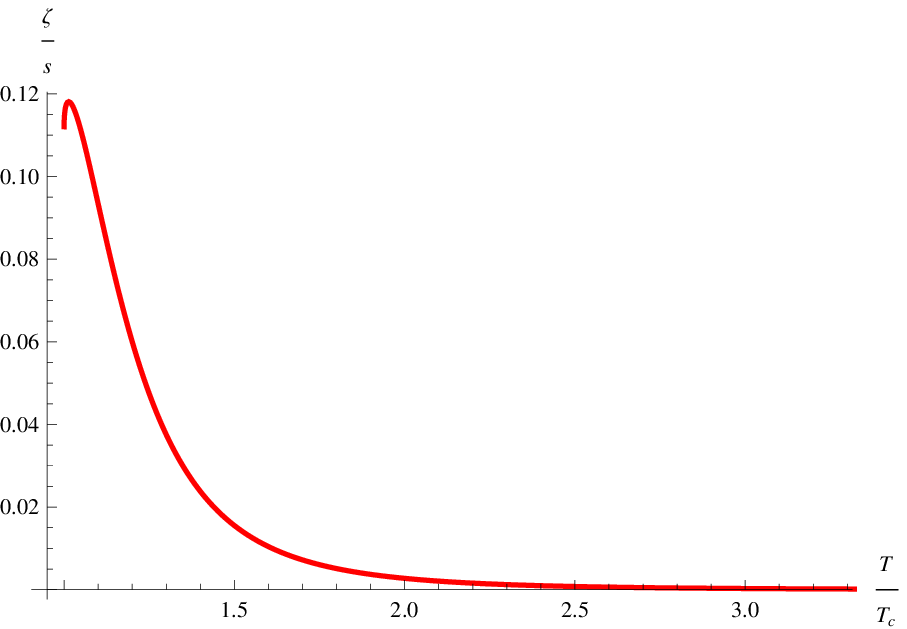} \hspace*{0.1cm}
\epsfxsize=6.5 cm \epsfysize=4.5 cm \epsfbox{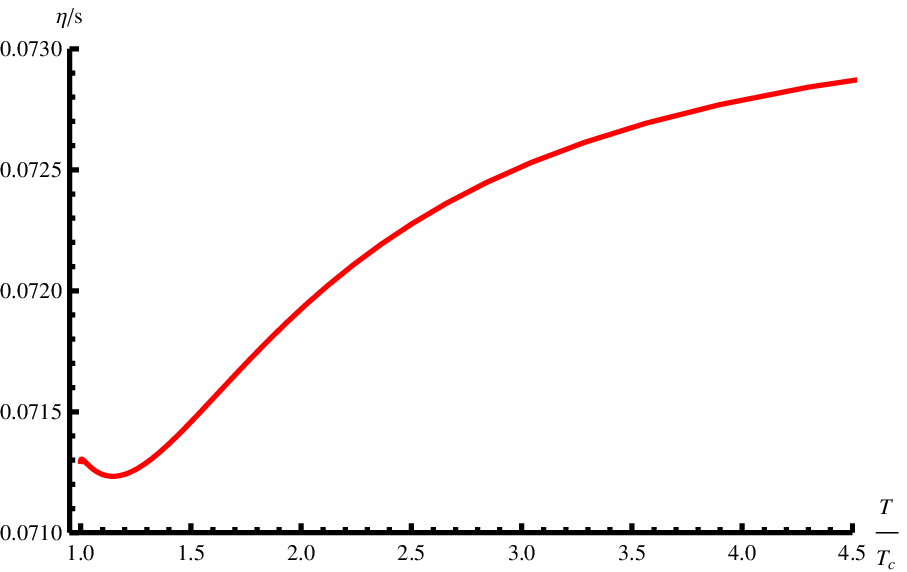} \vskip -0.05cm
\hskip 0.15 cm
\textbf{( a ) } \hskip 6.5 cm \textbf{( b )} \\
\end{center}
\caption[]{Bulk viscosity over entropy density and shear viscosity over entropy density
in the quenched dynamical QCD model. } \label{zetas}
\end{figure}

The bulk viscosity can be extracted by using the Kubo formula
\begin{equation}
\zeta=\frac{1}{9}\underset{\omega\rightarrow0}{\text{lim}}\frac{1}{\omega}\text{Im}\langle T_{xx}(\omega)T_{xx}(0)\rangle,
\end{equation}
where $T_{xx}$ is the $x-x$ component of the stress tensor.
According to the holographic dictionary, it's easy to extract the Green function of the stress tensor by considering
the corresponding metric perturbation $g_{\mu\nu}\rightarrow g_{\mu\nu}+h_{\mu\nu}$. Here, when applying the
Kubo formula, we have taken the spatial component of momentum $\overset{\rightarrow}{q}=0$ firstly, therefore,
we assume that $h_{\mu\nu}$ depends on $t, z$ only, i.e. $h_{\mu\nu}=h_{\mu\nu}(t,z)$. Explicitly, the corresponding
metric perturbation components should contain $h_{xx},h_{yy},h_{zz}$, and due to spatial rotational invariance,
one can choose three components to be the same. However, one can check that the $h_{tz},h_{tt},h_{zz}$ and the
dilaton perturbation $\delta \Phi$ are all coupled together, which makes the calculation quite complicated. Fortunately,
as pointed in \cite{Gubser:2008yx}, one can use the gauge degree of freedom to eliminate $h_{tz}$ and $\delta \Phi$,
 leaving only the diagonal perturbation. Besides, the authors also prove that the imaginal part of the retarded Green
 function $G_R$ is related to a conserved flux $\mathcal{F(\omega)}$ by the following equation
\begin{equation}
\text{Im}G_{R}=\frac{\mathcal{F(\omega)}}{4\pi G_5},
\end{equation}
where under the metric ansatz
\begin{equation}\label{metric-new}
ds^2=e^{2A}(-fdt^2+dx_idx^i)+\frac{e^{2B}}{f}dz^{2}
\end{equation}
and coordinate definition $\Phi(z)=z$, the conserved flux takes the form
\begin{equation}
\mathcal{F}(\omega)=\frac{e^{4A-B}f}{4A^{'2}}|Im h_{xx}^{*}h_{xx}^{'}|.
\end{equation}
More conveniently, the equation of motion for $h_{xx}$ component is simply independent
on $h_{zz},h_{tt}$ and takes the following form
\begin{equation}\label{hxx-eom}
h_{xx}^{''}=(-\frac{1}{3A^{'}}-4A^{'}+3B^{'}-\frac{f^{'}}{f})h_{xx}^{'}+(\frac{e^{-2A+2B}}{f^{2}}\omega^2+\frac{f^{'}}{6f A^{'}}h_{xx}).
\end{equation}
These results are quite general within graviton-dilaton system. Therefore, it's very easy to extract the bulk viscosity in our quenched dynamical
holographic QCD model by taking a new fifth coordinate as $z^{'}=\mu_G^2z^2$, which satisfies the requirement $\Phi(z^{'})=z^{'}$. Then we can derive the new metric form under the new coordinate
\begin{equation}
ds^2=e^{2A_s}(-fdt^2+dx_idx^i)+\frac{e^{2A_s}}{fZ^{'2}}dz^{'2}.
\end{equation}
with $Z(z)=\mu_G^2z^2$. Then it's easy to see $e^{2A}=e^{2A_s}$ and $e^{2B}=\frac{e^{2A_s}}{Z^{'2}}$.
Solving Eq.(\ref{hxx-eom}),  one can get the temperature behavior of the bulk viscosity as shown in Fig.\ref{zetas}(a)
with $\mu_G=0.75{\rm GeV}$ and $G_5=1.25$.
It is found that the bulk viscosity over entropy density shows a sharp peak near the transition temperature, and this feature
is in agreement with lattice results in \cite{LAT-xis-KT}.

For shear viscosity, any isotropic Einstein gravity system generates universal $\eta/s=1/4\pi$ result.
In order to introduce temperature dependence to shear viscosity, we follow \cite{Cremonini:2012ny}
and introduce higher derivative corrections of the following form
\begin{equation}
S=\frac{1}{16\pi G_5}\int d^5x \sqrt{-g}\big(R-\frac{4}{3}\partial_\mu\Phi\partial^\mu \Phi-V(\Phi)\nonumber\\
+\beta e^{\sqrt{2/3}\gamma\Phi}R_{\mu\nu\lambda\rho}R^{\mu\nu\lambda\rho}\big).
\end{equation}
Here the extra $\sqrt{2/3}$ factor in the coupling of $R_{\mu\nu\lambda\rho}R^{\mu\nu\lambda\rho}$ is to
keep our parameter $\gamma$ comparable to the one in \cite{Cremonini:2012ny}. As mentioned above,
to calculate shear viscosity, one has to consider the metric perturbation $h_{xy}$, and the shear viscosity
could be extracted through the Kubo formula
\begin{equation}
\eta=\underset{\omega\rightarrow0}{\text{lim}}\frac{1}{\omega}\text{Im}\langle T_{xy}(\omega)T_{xy}(0)\rangle.
\end{equation}
Up to $O(\beta)$, the shear viscosity over entropy density ratio results read
\begin{equation}
\eta/s=\frac{1}{4\pi}\left(1-\frac{\beta}{c_0}e^{\sqrt{2/3}\Phi_h}(1-\sqrt{2/3}\gamma z_h\Phi^{'}(z_h))\right),
\end{equation}
with $c_0=-z_h^5\partial_z\left((1-z^2/z_h^2)^2e^{2A}/(8f(z)z^2)\right)|_{z=z_h}$.
Fig.\ref{zetas}(b) shows the numerical result of  $\eta/s$ when $\beta=0.01,\gamma=-\sqrt{8/3}$, it is observed that
there is a valley at around $T=1.1T_c$, which is almost the location of the peak position in $\hat{q}/T^3$.

\section{Discussion and summary}
\label{sec-summary}

In this work, we construct a dynamical holographic QCD (hQCD) model
in the graviton-dilaton-scalar framework, where the dilaton background field
and scalar field are responsible for the gluodynamics and chiral dynamics, respectively.
The dynamical holographic model can resemble the renormalization group from ultraviolet
(UV) to infrared (IR). The metric structure at IR will be automatically deformed by the
nonperturbative gluon condensation and chiral condensation in the vacuum.
The produced scalar glueball spectra in the graviton-dilaton framework
and the light-flavor meson spectra generated in the
graviton-dilaton-scalar framework agree well with lattice data and experimental data.

The equation of state and transport properties are investigated in the dynamical holographic
pure gluon system. It is found that the trace anomaly $(\epsilon-3 p)/T^4$, the
ratio of the jet quenching parameter over cubic temperature ${\hat q}/T^3$ and the bulk viscosity
over entropy density show a peak around the critical temperature $T_c$, and the shear viscosity
over entropy density shows a valley around phase transition.

In summary, our dynamical holographic QCD model can describe hadron physics, QCD phase
transition, thermodynamical properties and transport properties quite successfully.

\begin{acknowledgement}
This work is supported by the NSFC under Grant Nos. 11275213,
11261130311(CRC 110 by DFG and NSFC), CAS key project KJCX2-EW-N01,
and Youth Innovation Promotion Association of CAS.
\end{acknowledgement}

\end{document}